# Course Material Selection Rubric for Creating Network Security Courses

By

Matthew Marriotti

This project is submitted to the Gannon University graduate faculty in partial fulfillment for the degree Master of Science in Computer and Information Science.

Option: Thesis (Plan A)

Approved:

Barry J. Brinkman, Ph.D.

Advising Professor in Charge of Research

Stephen T. Frezza, Ph.D., CSDP

Chair, Computer and Information Science Department

Theresa M. Vitolo, Ph.D.

Committee Member

John H. Coffman, MS

Committee Member

Michael J. Beiter, City of Erie School District

Committee Member

Gannon University

Erie, Pennsylvania 16541

December 2008

# Table of Contents

## Acknowledgements:

I would like to thank all the professors and friends that I have met at Gannon University. Times are never always easy, and having people to look to for support and guidance in difficult times is better than any grade. I would also like to thank my friend and mentor Ben Nagel, for allowing me to work with him in learning a great deal about networking techniques and how they are managed across large networks, and of course being able to just geek out over technology. Most of all I would like to thank my family for being supportive of my pursuits of higher education, especially my father Louis Marriotti who first introduced me to a computer at a young age. Allowing me to grow my interests with the subject to a point where I feel computers a part of my life. And even with the hardships that the computer might have brought us, I know it has brought us closer because of it.

## Abstract:

Teaching network security can be a difficult task for university teachers, especially for teachers at smaller universities where the course loads are more diverse. Creating a new course in network security requires investigation into multiple subject areas within the field and from multiple sources. This task can be daunting and overwhelming for teachers from smaller universities because of their requirement to teach multiple subjects, not just network security. Along with the requirement of teachers to understand the material that they wish to teach, the factors of obsolescence and the ability to build material off of core topics need to be addressed. These three factors are difficult for a smaller university teacher to address without a set of standards to analyze these areas. A rubric addressing these topic areas of timelessness, associability, and simplicity has been created to assist in the selection of materials based on the three criteria. The use of this rubric provides an effective means to choose material for a new course and help teachers to present the material they determine most appropriate to teach.

# List of Tables

## List of Acronyms

| | |
|---|---|
| AES | Advanced Encryption Standard |
| AIS | Automated Information Systems |
| IACE | Information Assurance Courseware Evaluation |
| INFOSEC | National Training Standards for Information Systems Security Professionals |
| IP | Internet Protocol |
| MU | Miami University |
| NS | Network Security |
| NSA | National Security Agency |
| OS | Operating System |
| PSU | Plymouth State University |
| TAS | Timelessness, Associability, and Simplicity |
| NSTISS | National Security Telecommunications and Information Security |

# Chapter 1 – Introduction to Teaching Network Security and the Rubric

Teaching network security (NS) provides an interesting challenge for universities. The size of the university is an influential factor in how these courses are laid out and taught. Larger schools (NS students numbering more than 50) generally have the means to provide not only the resources for the students to learn, but also the staff to keep the course fresh and up to date on NS topics. Smaller schools often do not have the numbers of students (although not necessarily less interested) enrolled in NS, but also the staff teaching the subject generally has other teaching obligations in addition to NS courses. This can make updating the courses more difficult for smaller schools due to time constraints, other obligations, and size of classes.

Compounding the issues regarding teaching NS is that material in this field generally reaches a point of obsolesce in a short time frame. Because of this faster paced progression, NS courses need to be updated frequently to reflect the changes in the material. Again, staff dedicated to NS courses can retool materials to fit these updated ideas, while smaller schools may be unable to continuously revamp course materials.

College courses in general are taught in similar manners year to year. For example, programming courses provide a background in the algorithmic methods used to solve problems, and build the actual language syntax upon solving the algorithms. NS does not necessarily have this luxury, due to the wide ranges of topics that can be presented. Schools can approach teaching the subject from multiple viewpoints; for example, teaching students how to program viruses or exploits vs. teaching the development of policy administration for networks. Both deal with NS, but in totally unique ways. The broad nature of the topic requires a different approach to teach, than say a programming course.

Developing a new course in general is not simple; NS is no exception, and in fact it may be more challenging than most, especially for smaller schools. In order to provide an adequate introduction to the subject matter, multiple topics must be addressed due to the scope of the field. However, because of the unique needs of smaller schools, ideas must be developed to address the matter of obsolescence. In addition to obsolescence, other aspects also must be addressed. The ability of the teacher to adequately present the material is one, and the ability of material to be built upon long lasting topics is another.

To accomplish the goal of assisting smaller schools to effectively and efficiency create course material which is easy to implement and able to stand up to obsolescence, a rubric for determining valid topics needs to be created. The formation of this rubric includes the idea of 'Timelessness'. Timelessness in terms of the NS rubric means the ability of a topic being able to withstand the ever occurring obsolescence in the field. For example, a security flaw in an operating system is not timeless, because, the speed a patch is applied renders the flaw neutralized in a short period of time. However, something like cryptology is very timeless, because the idea of cryptology, while changing, is an ever present part in NS. In a simple sense, something is more timeless the closer to abstract idea it becomes. This idea is where the core of a course needs to derive from. For the purposes of this rubric, the idea of timelessness applies to the material being presented; is the information presented at an abstract enough level in which changes to the field do not render the information taught useless?

A rubric cannot exist with one idea however; ‘Associability’ is a second area which is included. The idea of associability in terms of this paper, are topics which when placed upon the ‘timeless’ core material, brings the material into a current state of technology. For example, taking the idea of cryptology, and presenting the idea of the AES encryption method and its uses. AES one day will become obsolete, but the core of cryptology remains.  Generally associability and timelessness exist together, however conditions can exist where the ideas are excusive, for example, material which is highly associable, a in class activity for example, where the information can be changed depending on various conditions, can focus on a topic which is not timeless, such as a newly developed piece of software, which has not been around for a long period of time. For the purposes of the rubric, associability relates to the ability for the material to be altered to fit a new task or idea.

The final area which needs to exist for use in a smaller school NS course is simplicity. If a topic is not practical to teach, then it does no good to have a timeless and associable topic. Take the AES example; if the school does not address cryptology and the students are not versed in the material, for example, the school’s computer curriculum has a strong business element rather than programming. Then teaching the intricacies of AES encryption would not be as useful to students as perhaps teaching the management of systems for detection of security breaches. Also smaller schools’ staff may not be able to adequately teach certain subjects well. Simplicity for the purposes of the rubric compares the subject matter to the ease of integration and teaching in a college environment. These three areas of the rubric are described in further detail in Chapter 3 of this paper.

Currently ideas exist on proposed structures to present materials pertaining to NS. The government has had standards since at least 1994 when they created the National Training Standards for Information Systems Security Professionals (INFOSEC) (National Security Telecommunications and Information Systems Security, 1994). Other courses exist in both private and public sectors. All of the courses focus on particular areas of NS. Few have focused to a smaller university setting. Hurdles such as funding, class size, material determination all play important roles in how this type of course can be taught. The point of this paper is not to create a 'silver bullet' for teaching NS to smaller universities, but to present a means for these schools to take materials that already exist (textbooks, existing courses, standards, etc.) and choose and shape the materials to best fit their needs.

In order to construct a well developed and functional NS course for a small university, the ideas of timelessness, associability, and simplicity (TAS) are investigated and formed into a scale from which current works (academic and published materials) can be compared for best integration into  course. Based upon material which seems to fit the rubric of TAS, a sample NS course is created which can be integrated into existing computer curriculums to provide a strong base of knowledge for students to build from in their college careers. The rubric then can be used to investigate further materials and ideas, from which to easily integrate into the existing course. Over time, timeless materials remain relatively consistent, while the associable aspects can be swapped out to best teach new materials. This modularity allows for a relatively simple means from which to continue to provide up to date curriculum to students in a smaller university setting.

## Chapter 2 – Review of the Literature

In order to determine the best means for schools to approach the creation of a rubric based NS course, the existing subject materials relating to NS under these constraints need to be investigated to ascertain how best to form a guide for others to follow. Once the field has been investigated, it then is possible to form a rubric based on the criteria of TAS, to rate the materials and determine what would best be used to formulate a new NS course. Because a course requires a strong core, materials to support the core and supplies to support the materials, no one course or idea can form the entire basis or associable aspect of the final course.

The field of NS can be approached from multiple viewpoints, which all of the literature found does to one degree. In general there seems to be an acceptance of teaching the ideas and skills behind defensive techniques, as opposed to hacking techniques which seem to be viewed as a taboo subject by some. Each article addresses this information in its own way. Some papers and books embrace the idea of hacking; demonstrating various means to exploit for the purposes of introducing students to real world scenarios, which as professionals need to be addressed. These articles generally build upon an ethical basis to justify the teaching of this subject. Other articles choose to focus to defensive techniques without the teaching of the methods of exploitation. The observation regarding hacking is that professors that choose to present the material do so with the thought that simply presenting students with malicious ideas does not make the student necessarily a hacker. This argument is presented by some of the articles below.

Multiple angles and approaches exist to teaching NS; the articles below show this fact. Depending on the needs and desires of specific teachers preparing courses, some sources may be more useful than others. This collection of articles is by no means exhaustive or entirely comprehensive, but it does present a wide variety of courses and papers on the subject of NS, approaches to teaching the subject, and viewpoints on which aspects should and should not be addressed.

In 1994 the government developed a set of standards for training security professionals working on government computer systems, the National Security Telecommunications and Information System Security (NSTISS). The National Training Standard for Information Systems Security Professionals (INFOSEC) is one of six standards presented by the National Security Agency (NSA) for accreditation in the Information Assurance Courseware Evaluation (IACE) Program (National Security Telecommunications and Information Systems Security, 1994). This program can be taught by organizations, and upon completion, students obtain IACE accreditation. This particular document outlines the standards for information security professionals.

INFOSEC (first level of the six) presents the standard in seven pieces (listed below), each defined as either awareness, or performance level.

**Table 1: INFOSEC Certification Levels**

| National IA Education & Training Program (6 Certification Levels) | |
|---|---|
| NSTISSI-4011 | National Training Standard for Information Systems Security (INFOSEC) Professionals |
| CNSSI-4012 | National Information Assurance Training Standard for Senior Systems Managers (SSM) |
| NSTISSI-4013 | National Information Assurance Training Standard For System Administrators (SA) |
| NSTISSI-4014 | Information Assurance Training Standard for Information Systems Security Officers (ISSO) |
| NSTISSI-4015 | National Training Standard for Systems Certifiers (SC) |
| CNSSI-4016 | National Information Assurance Training Standard For Risk Analysts (RA) |

Awareness implies sensitivity (understanding) to the subject, and performance indicates the ability to perform the particular standard (National Security Telecommunications and Information Systems Security, 1994). These two main distinctions are used in the seven areas for teaching (5 awareness and 2 performance level). The distinctions are further refined into instructional and behavioral outcomes. Instructional gives a idea on what needs to be taught in a general sense, while the behavioral outcomes describe what the student should achieve in that section. The items taught to awareness level are broken into the following categories:

1. Communications basics, which describes the various types of transmission media, and the historical content of the field.
2. Automated Information Systems (AIS) describes software, hardware, and some additional networking ideas.
3. Security Basics, deals with information security and NS in general (cryptology, data integrity, etc).
4. NSTISS Basics deals with legal items, risk management, and trust (users trust to a “secure system”).
5. System Operating Environment, which deals with agency specific systems and security.

Performance level items include;

6. NSTISS Planning and Management, dealing with planning, risk management, contingency planning (creation of these types of documents).
7. NSTISS Policies and Procedures, describes the physical and non-physical security practices, both system and non system which students need to be able to perform.

The paper goes on to further describe some other areas of NS regarding trust, integrity, confidentiality, and availability, describing how they are important to NS and to the course layout.

The INFOSEC course structure focuses on information and not technology to provide a strong NS course structure, not teaching how to use items as much as relying on the theory and information to help prevent breaches in security. Of course a class taught with just information can become hard to follow, especially in the field of NS, with all of the abstract ideas. However the purpose of the INFOSEC paper is not to provide a course to follow to the letter, but a set of ideas upon which to build a course.

INFOSEC is part of an accreditation that schools can become certified to teach. Outside of the university, the base level of security certification is Security+. Security+ is offered by CompTIA, a developer of vendor-neutral IT certification exams. Security+ certification indicates mastery in security concepts and practices, or the equitant of two years of on the job training (Ciampa, 2004). Ciampa's book, *Security+ Guide to Network Security Fundamentals,* describes information required to pass the certification exam. Information is presented in a fairly easy to understand format, with exercises and activities to reinforce the chapters. The material in the book is given at a level where most of the material can be considered semi-timeless, or not fundamentally changing for multiple years. However some of the activities and labs presented are operating system specific, and at the very least, the steps required to perform the tasks can change (Most of the screens are presented from Windows XP).

This material is more in-depth than the INFOSEC standards because it is expected to be used as study material for a certification exam, and not just a set of standards to build courses upon. In some cases, material in the INFOSEC standard is drilled into fairly deeply in this book (e.g. ideas of trust and cryptology). This is mainly due to the changing security needs compared to the INFOSEC document of general standards. The book contains some practical labs which can be used with fairly little setup and equipment, which allows this book to be a good supplement for lecture.

One of the key areas of note within the scope of this paper is that the goal of the rubric is aimed at an organization which does not necessarily have the resources to provide a constantly updated and large network security course. Plymouth State University (PSU) is one such school. LeBlanc and Stiller, both professors at PSU, have developed a small one semester course in network security. The course they developed presents students with a good background in the area of security, teaching some core topics in lecture, combined with a set of labs which seem to relate to the lecture material (although it is hard to tell for sure due to just having a listing and not an outline). Lecture topics include pieces on threat assessment and Vulnerability awareness, with labs consisting of a mix of security and attacking tools (LeBlanc & Stiller, 2004). Students are also asked to author some papers, investigate security topics, and also write a security policy. These topics being taught relate to pieces of the INFOSEC standard, although no mention is given of the standards. The PSU course has the students performing multiple aspects of NS, especially in the hacking and defensive areas. This approach allows the students to have a more total experience in NS. The course also includes assignments for authoring some papers on subjects in the NS area. These papers are used to enhance the students understanding in an area of NS, and allow each student to provide material for other students to enhance their knowledge, without having a set classroom structure. This variability allows the course to teach subjects which were not planned to be taught, however it requires the students to take enough interest in the materials to present them in a manner which helps other students and themselves learn the material.

PSU is not the only smaller university which teaches a custom NS course. NS courses developed at smaller universities seem to be developed by the individuals who are going to teach the material. This means that the course created already has an expert in the material to teach. However, it also means that the course may be limited by the person(s) creating such courses, depending on knowledge and their desire to enhance that knowledge in creating a course.

Miami University (MU) also runs a course similar to PSU. Focused at the associate level, Laurie Werner used the materials from the NSA (authors of INFOSEC) in creating her NS course. Again the NSA has provided multiple course skeletons; it is just a matter of taking the ideas and building them out into a course. Similar to PSU, MU did not have a dedicated group of teachers to teach this program. One of the interesting aspects brought to the table at MU is the thinking that the hacking and administration aspects are similar, in that the tools administrators use; hackers do as well (Werner, 2004). Another key is that MU was trying to use information from other schools, and due to the nature of security, the presented papers and tools were outdated in some fashion, requiring some rework on her part. Each course found from smaller schools seems to bring references from the previous courses from similar sized schools. This idea is an important because it shows that others' work influence these courses more so than courses at a larger university, which have a more dedicated staff.

With courses like those at PSU and MU, there is a focus in teaching multiple aspects of NS, not only defensive tactics, but also offensive ones. Hacking is a subject which seems to be a hot-button issue in teaching NS because of its 'taboo' nature and the thought of possibly teaching the future criminals in colleges. Some schools take this idea and try to make hacking a legitimate part of an NS course, generally by introducing the idea of ethics in NS. This focus is what Brian Pashel a professor at Kennesaw State University has done in his course.

Pashel begins his paper by defining the three main types of computer hackers: white, grey, and black. He later describes why ethical hacking is such a difficult to understand subject. In general, ethical actions are ones that do not damage an individual or society. The main issue is that computers are not a thoroughly understood topic by the public, and therefore they do not understand the implications of hacking. Pashel simply defines ethical hacking as: hacking without malicious intent (Pashel, 2006). Ethical hacking is generally achieved in the same means as malicious hacking (Black Hat), but not with the entirely defensive minded ideas of preventing damage from attacks (White Hat), which makes an ethical hacker's hat, a fine shade of grey, where at times grey hats hacking, can be damaging or illegal, but it is not done with the intent to inflict damage. From these definitions it is easy to see why the ethics of teaching students the art of Black Hat hacking is a slippery slope indeed. However the main idea of Pashel's work is to stress that teaching students how to hack does not make them hackers, and the knowledge gained can be used to protect as well as for malicious intent.

Overall there does not seem to be any sources indicating that teaching defensive techniques is bad; all issues arising in teaching NS subject matter deal with the attacking side of the spectrum. There seems to be a divide in the matter. This divide is because the community as a whole seems to lean towards the idea that teaching and being are separate. The correlation between learning to hack and becoming a black hat hacker would be hard to prove. Just the same as the argument that rock and roll music of the 1970's degraded that era's youth. A simple cause and effect relationship does not exist between bad hackers and good ones. However, the hurdle of teaching this subject must be addressed from multiple angles, and not just from single sources. Dr. Logan and Allen Clarkson focus on the ethical side of hacking in their article, regarding the teaching of hacking in a university setting. This paper goes beyond Pashel's article, and describes a course layout in more detail. The paper does not layout a complete course however, but does discuss the idea of separating the lab experiments in a physically and digitally separate part of the university, allowing students to hack without the chance to infect the university's computer network. Even with this separate network, it is noted in the article that working closely with the information technology department of the university seems to be a good idea, to help calm the administration that the students are being properly supervised in these labs (Logan & Clarkson, 2005).

The main objective of teaching hacking is to allow students to see both sides of an attack, and perform security audits and unveil vulnerabilities in the systems they support, in short prepare students for the real world of network security administration. One of the main points made, which is cause for some concern over ethics in security, is that most universities do not require an ethics course in the computer curriculum, and students are simply taught a little bit about ethics in another class (Logan & Clarkson, 2005). Taking the time to present strong ethical course ties to the hacking and defensive teachings, help solidify the teaching, and also ease the concerns of individuals worried about teaching hacking.

Not all courses proposed by other schools have taken the hacking approach however. Mike O'Leary at Townson University has a group of courses that focus to network security, with the last of seven, *Case Studies in Computer Security* being outlined in his paper (O'Leary, 2006). This course is a hands-on lab based defensive minded course, which functions as a capstone in this school's security curriculum. O'leary states in his article, that attacking (Hacking) is not focused on because they wish to teach the defensive side of security, to *"teach potential security officers, rather than penetration testers"* (O'Leary, 2006). He further states that attacking is shown, but in less detail than the defenses. The paper goes into some detail of the labs being used to teach the course, all of which seem to be valid approaches to teaching defensive techniques. Also these labs fit well into the NSA standards, although the descriptions of the labs indicate that changes would need to be made over time to continue to teach the course. This change is almost unavoidable, but raises issues if the teachers are not familiar enough with the material to develop labs for the course as was the case with MU.

O'Leary does a good job pointing out counter points to teaching hacking in the classroom.  Again the overall idea of teaching or not teaching hacking seems to come partly from the instructor, but also from the environment in which the course is being taught. However if there was no counter argument to teaching hacking in NS, then there would be no argument at all. O'Leary brings a good counter point to the idea, and helps the argument overall because of it.

The above journal articles all deal with teaching security, with varying methods, but all share some commonalities, such as labs, activities, and lectures to engage students. These labs, activities, and lectures are a key groundwork for any course, especially courses which fit the constraints of being taught in smaller schools, with the need to develop a sound NS course. Not all useful sources are previously developed and published courses however. Many books exist that touch on multiple NS subject areas. While not all books are necessarily best suited for developing courses, many provide a valuable set of activities, and lecture supplements which can be very useful in a new course.

A book like *Computer Security*, authored by Matt Bishop is not a book that would be used for labs, but for lecture material. The content in Bishop's book is fairly abstract and because of this best suited to be used in explaining the ideas behind subjects. Material like cryptography and how to algebraically determine security is presented with great detail, however the key factor is, that the material is not tied to any software, or medium which can outdate (Bishop, 2002). The detail in the book makes it harder than many instructors would want to present in an introductory NS course, but the detail contained is invaluable for individuals who wish to know how NS works at a core level.

In contrast to Bishop's book, the *Hacking Exposed* series is a fine example of materials which can be used for activities and labs. The third edition of this book, deals with Windows Vista operating system hacks, along with hacks to Windows XP and Server 2000-2008 (Scrambray & McClure, 2008). This book focuses on circumventing security on the OS and is presented in a hacking manner, but also in a defensive mindset. This idea allows the book to function as a more complete learning guide, tying all the topics together so the student can see the results from multiple angles. This book would be a solid aid to labs because of its fresh nature, covering a newer operating system which remains usable for multiple years. Windows was chosen over Linux or Mac because of the prevalence in the work place, it can be assumed that most students have more familiarity with Windows. This book has been published for multiple editions, generally following each iteration of the Windows OS. Therefore it is important to find the book which most closely matches the material being presented.

Similar to *Hacking Exposed, Computer Security and Penetration Testing* focuses on the attacking side of security. The book takes a similar approach to the Security+ book, in that it presents topics, activities, labs, and screen shots of programs being used (Basta & Halton, 2008).The sections on password cracking, sniffing, and social engineering are good chapters which can provide labs/activities as well as lecture material. Some sections are already showing some age, such as IP spoofing, because of the focus on IPv4. Even with some outdated materials however, most of the information is useful and can be used to either build upon some ideas in Security+ or on the INFOSEC standard itself. A good mix of timeless material exists, along with some material which becomes outdated slowly, making this book a good resource for multiple classes.

While published articles on teaching NS provide a good set of examples of previously created content, books provide additional information from which courses are built. The above list is not exhaustive of the sources used in this paper, but the listed articles and books are a strong foundation for the created rubric. In Chapter 3 there is further investigation of book sources as the rubric process is further described and investigated.

# Chapter 3: Understanding the Rubric and Course Creation

## 3.1 Introduction

The studied material on NS shows that there are not only multiple varied sources of material, but multiple ways to approach teaching them. In order to teach a subject such as NS, there must be some criteria for choosing topics to teach. To build out the criteria to determine material to include, the most important aspect to consider is the end user of this information; a professor at a small university designing class. The criteria to determine which material to include into a course must take into account not only the teacher of the subject, but also the material itself. After investigating the materials and abilities of the teacher, a more effective course can be created which is best suited to the teacher.

Limiting factors faced by teachers in creating new courses are: Time, the available time for the teacher to create a new course. Instructor ability is the current or accessible knowledge of the subject, which is to be taught in the class. And budget, the resources available to the teacher. The three limiting factors for selecting and teaching material can vary from school to school. This area is very dependent on not only the school environment, but also the person teaching the material, his/her skills, time, and willingness. NS is fairly unique in that there are multiple areas to teach, and generally the material taught depends on the teacher and how he/she structures his/her course. For example, some may teach NS as a programming heavy course (writing malware or other networking programs). This type of course can also be taught from a more administrative or preventative nature. This is evident in the aforementioned articles, where each class had uniqueness depending on the teacher. Because these aspects of choosing a course are unique to the environment in which the teacher resides, the above ideas (time, ability, and budget) are rolled into one overarching term called “Simplicity”. This idea is a broad term for how able a teacher is to teach the course, not only in the teachers’ skill set, but also in the ability to obtain materials and funding for his/her course.

Simplicity is not the only idea that must be considered. The environmental (outside) factors of NS must also be taken into account when developing a course. The environmental factors of NS include the speed of obsolescence of the materials, and how those materials can be elaborated upon for teaching. The obsolescence of material is of great importance to a teacher of a small university when it comes to creating material for a new course. Depending on the length of time between updating the course, material the teacher has chosen may become obsolete and useless to present. Teaching items like operating system (OS) vulnerabilities may be very important, but once the issue has been handled by software providers, teaching about the problem does not apply as heavily as if the vulnerability still existed. The same idea applies to teaching about an OS which is at the end of the life cycle. For example, it does little good to teach earlier versions of Windows, especially in this environment (NS), where security issues are found and fixed in a short period of time. How does one determine what material is going to out date? An assumption can be made, that at some point, all material becomes obsolete, but choosing the materials which fit well into the smaller university NS course for a long period of time, a concept which is referred to as “Timelessness” or how well a subject stands up to obsolescence, or a topic which is abstract in nature, and can be considered a core of a subject. An example of this could be cryptology.

At this point there are two main ideas from which to choose material for courses: Simplicity, and Timelessness. Finally there must be some way to look at a subject and determine if the item can be adapted for teaching in the course. Being able to build upon topics and present activities and labs based on that subject in an easy to implement means. Because teachers need to reinforce the teaching with activities it is important to know that the material she/he decides to teach is able to be built upon easily in order to provide a strong backing for the subject core. The idea behind this is that teaching an abstract idea is helped by allowing students to experience the idea in an activity to help reinforce the idea. This thought is fairly separate from the previous idea of Timelessness; however this idea is not necessarily exclusive of it. The ability to build upon a core subject in this way is referred to as "Associability". More simply, associability is how flexible the selected material is in terms of its ability to be built upon and from. Associability also can refer to the ability for the subject to be removed from the course and replaced in a manner which does not hurt the area that it is being pulled from. This idea is why a timeless item can be associable as well, generally a core topic can be built upon easily, or it is easy to relate to its root item (the core/timeless topic). For example, cryptology is both timeless and associable, because it is a core topic, and can be built upon by becoming more detailed on the idea of the core of cryptology. While something like IP Headers is sort of timeless (changes to the IP header over time) but not very associable, because once you go into the detail of the headers, it is much harder to build additional content from it. Associability is how easily a topic can be built upon and derived back to. Topics which score well in both timelessness and associability (like cryptology for example) are in a unique position in that it can be used as either an associable or timeless topic. There should be a distinction made and the subject treated as either. This distinction means that while cryptology meets both criteria, if it is chosen to be a timeless topic, then material is built upon it as a core topic, while if it is chosen as an associable topic, then it can be built upon another core topic, perhaps networking for example.

Timelessness, Associability, and Simplicity are the three areas which can help teachers determine material to choose for his/her NS courses. These items are based upon a smaller school; however they can easily be applied to larger schools as well. Simplicity can have a broader meaning in larger schools, where resources (teachers, money, students) are generally higher, and because of this, simplicity does not apply as well but can be augmented if needed. Also the other two areas can be further broken down to span multiple courses and classes. The idea is to create a concise course for a smaller curriculum which may only have time for one NS course. This thesis presents the use of these three criteria, to best make use of the time and resources available to the teacher. However these ideas presented can be easily scaled to support larger curriculums.

## 3.2 Rubric

From these three ideas, a rubric has been created to help teachers decide on how to best make use of their time and energies for teaching. Like most rubrics each row is an individual score to assess the Timelessness, Associability, and Simplicity of a topic. While Simplicity should be high for all topics because of the need for the ability to reasonably teach a topic, timelessness and associability may not be necessarily rated as high for a topic. This occurs because some topics are best suited for cores of the course (timeless topics), or subjects which are abstract in nature, while associable topics are very detailed parts of a core (timeless) topic, and therefore may not be timeless due to their less abstract nature. These detailed ideas generally score high on associability, and hopefully timelessness. Topics which are less timeless should be used for labs and activities due to their higher chance of obsolescence, but generally they are more associable since they are built upon a more abstract (timeless) topic. If topics are low on timelessness and associability then they need to be carefully considered before including, due to the risk of having to either update the material quickly, or because they are "dead end" items, where not much can be built upon the idea. This rubric is not as cut and dry as some, where all "5's" is best. This rubric is to be used to determine the usefulness of topics, and to help decide on their inclusion or exclusion into the course being created. For example a subject that is highly timeless but not associable is not necessarily worse than a topic which is timeless and associable; it is up to the teacher to decide which particular criteria is necessary for a particular topic. Multiple dimensions are presented in each area of the rubric, where perspective topics can fit in more than one box on a rubric topic. If these situations arise, the average of the two areas can be taken unless one dimension is desired to carry more weight overall, then this consideration should be taken by the teacher.

**Table 2: Timeless, Associability, & Simplicity Rubric**

| Idea | 1 – Lacking | 2 | 3 | 4 | 5 – Mastery |
|---|---|---|---|---|---|
| **Timelessness** – Ability to withstand the obsolescence of ideas over time. | Idea or concept lacks any abstraction, or makes references to material which is specific to a unique instance which is dependent on time. | Idea or concept contains multiple details of which if any one becomes useless, the idea cannot be used without rework. | Idea or concept contains some detail and some abstraction, and will out date over time, but can be augmented to lessen impact. | Idea or concept is mostly abstract; examples are presented without referring to excessive number of changing ideas or concepts. | Idea or concept contains no specifics to topics and is abstract. Can be considered a root from which other ideas derive. |
| **Associability** – Ability to build upon or from a core (timeless) idea. | Idea or concept is unable to be related to a core element directly or indirectly, and/or lacks ability to be built upon or from. | Idea or concept cannot be directly related to a core element, but can indirectly be linked. Building upon the idea requires determining the core idea thru indirect means. | Idea or concept can be linked to core element directly or indirectly, but may not easily be built upon. | Idea or concept can be linked to core element directly. Additional content can be built upon with relative ease. | Idea or concept is distinctly built from a core idea which can be determined directly. Additional ideas can easily be built upon. |
| **Simplicity** – Ability for material to be taught by staff effectively and efficiently. | Idea or concept is out of skill set of teacher, or unable to be easily implemented with given resources or obtained resources. | Idea or concept can be learned by teacher with effort, and/or outside resources need to be used to great extent. | Idea or concept can be learned or taught in tandem. Resources may be needed but can be obtained with ease. | Idea or concept is known by teacher with little need for further understanding. Resources are within easy access. | Idea or concept is able to be taught and implemented without any outside influence on teacher. |

### 3.3 Using the Rubric to Choose Course Material

The rubric is designed to be a guide to assist in the choosing of topics to include in an NS course and its use is dependent on the person using the rubric; depending on the knowledge and comfort of one teacher, some subjects may be considered unsuitable for inclusion, whereas another teacher may find the same material useful. This difference is because the rubric is designed to assist in the choosing of material, not to provide a list of all material to use. However the rubric can be used by all individuals who are interested in developing a course; it simply provides a means to compare materials to an individual's skill set, resources, and needs of his/her course.

The three sections of the rubric require the teacher to analyze the possible topics chosen. This process allows for a more analytical view of the topics, especially as to how they might apply to the constraints of a smaller school (teaching time, resources, and knowledge level). However, the rubric allows for a personal touch to choose topics that a teacher is more interested in teaching. This uniqueness makes the rubric adaptable to all environments, as the rubric does not determine the material, but enables a more analytical look at the material and how it can be best used for teaching.

#### The Criterion

Simplicity implies the perceived ability to successfully teach a topic. This criterion can be considered the most important of the three due to the direct relation to the teacher of the material and is/her ability to present the material. This score should be an honest representation of the ability to successfully relay the information to the students, and an analysis of the economic or political difficulty in acquiring the material to be presented. Having the ability to learn or present the material in tandem with teaching can allow for topics to be included when they would otherwise be unable to be taught. However, consideration should be given to these topics and how the students are impacted by this method. An ideal topic would be understood well enough by the teacher to present the topic without the need of learning of the idea while teaching. Also the material needs to be easy to implement in the monetary sense. A topic which is well understood by the teacher but unable to be implemented within the budgetary confines of the university is less likely to be included in a course than a topic of which the teacher has less overall knowledge but easy access to.

An important fact about the rubric which should be noted is that certain topics in NS may not be easy to teach or implement by a teacher, but may be very important to NS in general. Because the simplicity factor of the rubric takes into consideration the unique situations of the teacher, it may appear that important topics are unsuitable for inclusion. The rubric is not a final word on inclusion or exclusion of materials, only a means to analyze materials compared to the instructor's ability to teach them. If anything, a perceived important topic with a low simplicity is shown to require more effort on the part of the teacher, and if he/she wishes to include the item, they can understand that additional time and resources may be required prior to teaching that particular topic.

Timelessness denotes a material's ability to withstand the obsolescence over time. For the purposes of developing a new course, this idea should be applied to how close to a “core” the topic is. Building a course in this field requires a set of topics from which to base lectures and labs upon. Rating a topic as “timeless” means that the idea itself, is as close to an abstract idea as possible, this allows for the topic to be built upon and from for lecture and lab. Also being abstract allows the idea to not be as affected by the changes of the NS field. Building a core upon abstract ideas alone does not make for an interesting class; however the purpose of this criterion is to find material which can be reused and built from over time. For the example of the course being built in this paper, the highest level main topics are Administration, Hacking, and Defense. From these three main core topics, the less abstract timeless ideas are applied to build the main objectives out, all the way down to labs and activities based upon non-abstract ideas.

Associability relates to timelessness in that an associable topic is built from a timeless topic; however topics can be less timeless and remain associable. An example is operating systems (OS); OS’s are not very timeless in a computer sense, with updated versions coming out with various frequencies, however OS’s are a fairly associable topic in that they are easily built upon for teaching. A second factor in associability for the rubric is how easily a topic is related to a core topic. This relation is important for when the associable topic outdates and needs to be replaced by new material. Knowing the root of the topic is important for replacing the idea with a new one. Using the OS example, having lessons on Windows XP system administration, is easy to relate back to its OS core (Windows XP), thus being an associable topic for its ease of updating to a more current OS. A topic which is very detailed such as the details of an IPv4 header (IPv4 Source address for example) ranks lower on associability, one because it is less able to be built from (detailing the IPv4 header is about as deep as one can go) but also because it is not directly related to the highest root level from which IPv4 is described, the OSI model. IPv4 headers need to be backed out though multiple topics to indirectly arrive back at the OSI model. While IPv6 is a logical transition from IPv4 in terms of teaching, this makes the root of IPv4 (IP in general) a more associable topic because of the ability to easily transition the whole section to a newer version. However IP is not the most abstract that this section could be, therefore while IP headers relate back to IP in general, IP itself must relate back further to the OSI model, making IP headers an indirect link to the root. The number of topics from the root topic is a consideration for scoring a topic as associable.

### 3.4 Sample Course Using Rubric

To best demonstrate the use of the rubric in determining material to present in an NS course, is a basic 15 week course is developed using the rubric as a means to determine content to include. The course is developed focusing on a combination of materials researched in the area of smaller university NS courses. The focus of the course focuses on a trio of topics: Administration, Hacking, and Defensive techniques (AHD). These three main topics were chosen after doing preliminary research on the topic of NS as a whole, and determining that the particular course would focus on these topics at a high level. Keeping in mind the criteria of the rubric while doing this preliminary search, choosing the main ideas to focus to, allows for the selection of materials based upon both the rubric, and the overall interest level of the teacher. Once these core ideas are picked, a set of course objectives can be created.

**Course Objectives:**

1. Provide students an overview of NS, the current state of the field, and the ongoing improvements in the areas of Administration, Hacking, and Defense
2. Introduce the areas of Hacking and Defense, providing insight into how hackers work, and how to defend against these techniques
3. Present administrative techniques to help prevent hacking, and improve defense of a network

From these objectives, a list of topics is compiled. The rubric is then applied to the list. This process allows the teacher to list the overarching ideas to cover, and then determine which of the topics are feasible to pursue further.

The decision to teach hacking methods in an NS course is totally up to the individual teaching the material, and whether they can safely contain the possible problems that arise with teaching possible volatile issues. Logan presents a good article on the implementations of teaching this subject matter, and possible issues that can arise with doing so (Logan & Clarkson, 2005). The article provides a strong set of ideas to address before implementing the course, in terms of hardening the classroom to possible issues, which should be understood before developing a hacking element of a created course. Some do not believe that teaching hacking in schools is any worse than teaching other possible dangerous activities to students. Because hacking is an important aspect of NS, a strong foundation in understanding the malicious side to hacking is important to teaching means to prevent those attacks. The individuals who prepared NS courses seem to embrace this idea, but not all do. Ultimately the existing curriculum and the teacher's ability should be the means to determine if teaching hacking is justified.

The list below is a set of topics from which the sample course is created. At this point in course creation, listing the topics is more important than the details. A long list allows for the application of the rubric to determine what is feasible to attempt to teach. For pass one of the list of possible topics, focusing on the "Simplicity" factor shows which topics are most easily taught by the teacher. Below is a sample of applying this idea to a list of topics for the sample course. These topics can be found one of three ways: one method is to use the INFOSEC text to find a list of high level topics that might be addressed, and finding their current implementation. For example, INFOSEC makes reference to teaching "Crypto security" which is referred to as cryptology in most texts, or "software security" which would boil up into the more current OS's and specific security with each (National Security Telecommunications and Information Systems Security, 1994). This method would allow for the most abstraction, but may be harder to relate all the listed material to more current references. A second method would be to generate a possible topic list from multiple text sources, keeping in mind the relation back to the core topics that the material derives from. A third method which is described more in Chapter 4 would be to use existing courses to help choose material. The length of the course and the constraints of the smaller school setting have an impact of the overall choice of course material. This factor is partly reflected in the simplicity factor of the material.

Whichever method listed above is the one(s) chosen to begin to pull material from, it is important not to limit yourself at this point to any number of topics. Over the iterations of applying the rubric, unfit topics can be removed, leaving a list of topics which are suitable for teaching one or more courses. This phase is one of the most important in selecting material, because while the rubric allows for analysis of topics based upon the unique needs of the teacher, the rubric does not indicate if topics have been missed from inclusion. Because this rubric is not a “black box” of creating courses, it is required that the teacher using the rubric understand that the more sources and raw topics that are analyzed with the rubric up front, more likely the end result should contain a more well rounded selection of good topics.

**Table 3: Possible Topic List Pass 1: Simplicity Rubric**

| TOPIC | Simplicity |
|---|---|
| Hacker Profiles | 4 |
| Goals of network security | 4 |
| Process of attack | 3 |
| Local server port control | 3 |
| Basic network | 5 |
| Social Engineering | 4 |
| Port sniffing /Packet sniffing | 4 |
| Proxies / basic network attacks | 4 |
| DDoS (Distributed Denial of Service Attacks) | 2 |
| Buffer overflows | 2 |
| MAC Spoofing | 3 |
| Access control lists | 3 |
| Network | 5 |
| Wireless security | 4 |
| Password Cracking | 4 |
| System Vulnerabilities | 4 |
| Risk assessment | 4 |
| Disaster Recovery | 4 |
| Encryption | 4 |
| IDS (intrusion detection) | 2 |
| Hacking Scanning tools | 3 |
| TCP Packets / Headers | 3 |
| TCP/IP Vulnerabilities | 3 |
| Spoofing | 3 |
| SMTP Vulnerabilities | 3 |
| Web Vulnerabilities | 3 |
| Proxies | 3 |
| Sniffing | 3 |

The above topics were chosen using a mix of two possible methods of topic selection. The main method used above for deciding upon which topics to address is to simply browse a selection of text books for topics which score higher on the simplicity scale (a rough idea of how capable you as a teacher feel on addressing the topics). At this point, rolling simplicity into the list allows the teacher to start to access his/her skill set, but at this phase it is important to keep all topics in the list, even if they would score low on simplicity. This overflow is done in case additional topics need to be used, or if low ranked topics are found in multiple sources, it may be useful to learn this particular subject. Keeping in mind that if certain topics appear in multiple sources (cryptology and encryption are examples using the above list) that there may be a reason to make attempts to include the material at some level (the depth of coverage into the subject will depend on the rated simplicity of the topic).

The second area that influenced the above topics was the INFOSEC standard. The information in this source is at a very abstract level, and therefore provides a good skeleton of a course that topic ideas can be chosen from.

At this point there is a general understanding of what topics can easily be included, and which require more work on the teacher's side, or which must be dropped due to difficulty of implementation.

Once the material chosen has been ranked according to the teacher's perceived ability to teach the material, applying the rest of the rubric can be done at the same time. These two additional criteria can be determined from one of two means; teacher knowledge, or research. Depending upon the simplicity ranking by a teacher, he/she may assign values depending on his/her perceived understanding of a subject. Otherwise performing research on the topic areas (finding material regarding the topic in existing books or articles) can shed light on whether a subject is timeless or associable. There is not absolute necessity that these criteria be applied with excessive backing; the values provide a means for a teacher to decide how the material can be applied to his/her course. Below is a sample of the example course and the scores assigned assuming the author would be teaching the material. The criteria are based upon using possible text books which contain the topics. Notice how the more abstract an idea, the more timeless it becomes, but this relation does not necessarily mean that the topic is not associable; there is not an exact 1 to 1 ratio.

**Table 4: Possible Topic List Pass 2: Including Timelessness and Associability**

| TOPIC | Simplicity | Timelessness | Associability |
|---|---|---|---|
| Hacker Profiles | 4 | 4 | 2 |
| Goals of network security | 4 | 4 | 4 |
| Process of attack | 3 | 4 | 3 |
| Local server port control | 3 | 3 | 3 |
| Basic network | 5 | 4 | 4 |
| Social Engineering | 4 | 5 | 5 |
| Port sniffing /Packet sniffing | 4 | 3 | 4 |
| Proxies / basic network attacks | 4 | 4 | 3 |
| DDoS (Distributed Denial of Service Attacks) | 2 | 3 | 2 |
| Buffer overflows | 2 | 3 | 3 |
| MAC Spoofing | 3 | 3 | 3 |
| Access control lists | 3 | 3 | 3 |
| Network | 5 | 5 | 4 |
| Wireless security | 4 | 4 | 4 |
| Password Cracking | 4 | 4 | 4 |
| System Vulnerabilities | 4 | 4 | 4 |
| Risk assessment | 4 | 5 | 4 |
| Disaster Recovery | 4 | 5 | 4 |
| Encryption | 4 | 5 | 4 |
| IDS (intrusion detection) | 2 | 3 | 4 |
| Hacking Scanning tools | 3 | 4 | 4 |
| TCP Packets / Headers | 3 | 3 | 3 |
| TCP/IP Vulnerabilities | 3 | 3 | 4 |
| Spoofing | 3 | 3 | 3 |
| SMTP Vulnerabilities | 3 | 2 | 3 |
| Web Vulnerabilities | 3 | 3 | 4 |
| Proxies | 3 | 3 | 3 |
| Sniffing | 3 | 3 | 3 |

Once the possible materials have been ranked according to the TAS rubric, the topics which score low in simplicity (1 or 2) should be filtered out, due to the difficulty in implementing these ideas. However it may be necessary to pull from these topics if a lack of material exists, or certain core topics need reinforced.

After filtering out the topics which are difficult to implement, the remaining topics should be investigated regarding the timelessness and associable aspects in regard to simplicity (high simplicity compared to the timeless and associability). Take from the list enough topics to teach over a semester course (varying depending on the length of the course). In this example course, administration, hacking, and defense is the main core, and the additional topics are drawn to reflect theses main topics. These core topics should rank highly in timelessness because these topics are to be used for multiple semester classes. If the topics also have higher scores in associability, it is easier to build from these topics because of their nature; however it is not a necessity for the core topics.

The remaining topics need to be viewed from an associability aspect. These remaining topics are what can be used to build out the activities, labs, and augment the lectures based on the core topics. These topics need to rank higher in associability, and should be able to be related to the core topic easily. This area is where topics can be focused more to the current state of the field of NS. Of course topics which are fairly timeless should be used first and form the more important activities and labs which can be a “core” or associable ideas. However topics for this section can be ones which out date quickly and even be topics off the list of the proposed ideas. The list of ideas is more for developing the timeless and simple topics to teach. Any topics chosen not on the original list should be compared to the rubric in terms of the ease of implementation and its associability.

**Table 5: Possible Topic List Part 3: Sorted Data with TAS**

| TOPIC | Simplicity | Timelessness | Associability |
|---|---|---|---|
| Network | 5 | 5 | 4 |
| Basic network | 5 | 4 | 4 |
| Social Engineering | 4 | 5 | 5 |
| Risk assessment | 4 | 5 | 4 |
| Disaster Recovery | 4 | 5 | 4 |
| Encryption | 4 | 5 | 4 |
| Goals of network security | 4 | 4 | 4 |
| Wireless security | 4 | 4 | 4 |
| Password Cracking | 4 | 4 | 4 |
| System Vulnerabilities | 4 | 3 | 4 |
| Proxies / basic network attacks | 4 | 4 | 3 |
| Hacker Profiles | 4 | 4 | 2 |
| Port sniffing /Packet sniffing | 4 | 3 | 4 |
| MAC Spoofing | 4 | 3 | 3 |
| Hacking Scanning tools | 3 | 4 | 4 |
| Process of attack | 3 | 4 | 3 |
| TCP/IP Vulnerabilities | 3 | 3 | 4 |
| Web Vulnerabilities | 3 | 3 | 4 |
| Local server port control | 3 | 3 | 3 |
| Access control lists | 3 | 3 | 3 |
| TCP Packets / Headers | 3 | 3 | 3 |
| Spoofing | 3 | 3 | 3 |
| Proxies | 3 | 3 | 3 |
| Sniffing | 3 | 3 | 3 |
| SMTP Vulnerabilities | 3 | 2 | 3 |
| IDS (intrusion detection) | 2 | 3 | 4 |
| Buffer overflows | 2 | 3 | 3 |
| Denial of Service Attacks | 2 | 3 | 2 |

Now that the list of topics is sorted along the criteria listed above, depending upon the length of course the topics need to be taken and molded into a structure to be taught from. The example course is 15 weeks long, and based on the topics of Administration, Hacking, and Defense. These topics have the ability to share subtopics among them. For example, the timeless and associable topic of social engineering has topics that can be taught in all three of those main cores. Because of the length of course, and the three main areas that want to be addressed, the top 10 topics are taken and further developed into a course. These topics rank highly in all three of the rubric criteria. The topics below the top ten can be supplements to these main topics and be further refined to fit into the main ideas being taught.

**Sample Course Selected Main Topics and Reason for Choice:**

When detailing out the material chosen, it may be useful to cite the primary source for the material if one source is used more than others. As shown below, some items are focused more to a particular source; this source is the main article containing the lecture/lab material.

1. **Basic Network**

This topic is the main overarching topic for NS courses. A basic overview of the mechanisms for transmission of data should be covered prior to investigating deeper into the subject. This section consists of the basic setup of networks to reinforce the lecture, but also to prepare the labs for the other activities. The idea is to have the students understand the basics of how hacking affects networks as a whole. This subject is timeless because of its abstraction and also associable because, all topics extend from this topic in some manner. Because the existing framework should exist to teach networking, and the teacher should have to have an understanding of the subject to consider teaching a course of this type.

2. **Social Engineering** (Simon, 2002)

Social Engineering contains both computerized and traditional methods of gaining information from individuals. This subject deals with the trust in the area of security in general especially with the trust of the human element of security (Simon, 2002). Case studies can serve as activities linking administrative techniques of preventing such attacks. Hands on social engineering can demonstrate how this method can be implemented by hackers. The topic is timeless in that this technique has existed essentially forever (computer and non computer). Social engineering is fairly associable with the ability to further break the idea into defensive and offensive techniques, and relates to trust directly.

3. **Risk Assessment** (Holden, Guide To Network Defense and Countermeasures, 2003)

    Risk Assessment is mostly be taught from the administration side of the course, laying out the approaches to securing data both physically and digitally. The idea of survivable network analysis should be investigated to show both defense and attacking areas that need to be investigated to maintain the core idea of trust in NS. This idea is very timeless due to its need in any environment where maintaining trust is needed. The idea behind risk assessment has no software requirement to out date, and can continue to evolve with new and emerging technologies. A high level of associability exists with this subject because of its ability to be branched off and investigated from all three of the core area of the NS course. The idea also relates to trust in a direct manner which is the overarching focus of the INFOSEC course skeleton.

4. **Disaster Recovery** (Ciampa, 2004)

    Disaster recovery is another section to be taught from the administrative side of the course. Building understanding for how to recover from an attack or disaster should be the focus of this section. The section is built from risk assessment, and should include developing a disaster plan and securing systems and backups. This section is timeless much like risk assessment is. Wherever data exists

that requires protection, disaster recovery exists in some form. Disaster recovery's associability is also high because it builds from risk assessment and the teaching of the material can come in multiple forms depending on how the information is going to be presented.

5. **Encryption** (Bishop, 2002)

   Encryption is a section which can be based on the idea of how cryptography works, and how it is implemented in today's world. The timelessness of the topic comes from the theory of this subject, with the associability coming from investigating how the ideas are currently being implemented. Due to the difficulties in breaking strong encryption, this section should take a more defensive tone along with an administrative one. Simple encryption breaking can take place, along with activities in hiding data.

6. **Password Cracking** (Basta & Halton, 2008)

   Password cracking is a fairly basic idea in the hacking world. For a smaller course it would be useful to keep the idea at that level, and introduce the ideas behind the technique. Showing some existing programs that make this process easier would be a good activity, as would teaching the tactics of social engineering and guessing common passwords. This idea is rather timeless in that passwords generally are a weak link due to people, and until passwords are fundamentally changed they will remain a high risk. This topic is associable in that it is fairly easy to go in multiple directions from these topics (software, social engineering, etc). This idea relates to trust but not directly, but because of its ease of building upon it is more associable than not.

7. **Wireless Security**

Because wireless networks have become popular in both businesses and home use, it would be a good topic to introduce in this class. The focus will be on topics of methods of securing (both digitally and location wise). If possible, running activities with some wireless encryption crackers would be a good activity, along with circumventing some common methods for securing networks. This topic is fairly timeless in that wireless is fairly prevalent in the world today, but it is fairly young in terms of lasting technology. The topic is associable however, in that multiple teaching angles can be applied to this topic, depending on interest and the knowledge of the teacher to expand upon this topic, possibly including some Linux for the OS's strong suite of wireless applications.

8. **System Vulnerabilities**

System vulnerabilities present a good way to integrate existing problems with computers into the course. This section is associable in that it is easy to find and build activities that show system vulnerabilities; however the vulnerabilities themselves are not very timeless. A topic such as this would best be suited to focus to a different core area, such as network security and draw the activities from those areas that relate back to system vulnerabilities. Due to the nature of this area, it is best to pull examples as needed from more current resources and not books, unless those books are going to be used as case studies of more large scale vulnerabilities.

9. **Proxies / Basic Network Attacks**

   This section deals with some basic networking attacks, such as man in the middle, some basic network hardware attacks, and denial of service attacks. This section can be used for basics of “hacking” to demonstrate how they happen on ill prepared systems. Proxies can be used to show how hackers keep anonymity while performing attacks. This area of study can easily be related to the three core areas, and thus is very associable. Depending on the attacks chosen, an element of timelessness can exist, although the means to accomplish the attacks may change over time.

10. **Hacker Profiles** (Basta & Halton, 2008)

    The profile of hackers provides a good background for how certain types of hackers think. This understanding is important to administrators and defensive security individuals in developing countermeasures for hackers. This topic is fairly timeless because the main archetypes of hackers have changed little in the past 20 years. From the different types of hackers, activities and case studies can be used to see how each type of hacker influences the networks that he/she attacks.

The above list is a sample of how the rubric would be used to choose topics, and the justification of these topics to further build out a course. Detailing the reasons behind choosing all materials, allows the teacher to put her/his thoughts down to further help in fleshing out ideas. At this point in the course building, a basic curriculum would be created outlining the material which would be taught each week. Because this paper is not focusing on presenting a course but rather on a method for developing a course, the below base curriculum shows the main topics, and how the main three core topics can be addressed from each of the ten main topics. Further specifics would be built out as the course is created by the teacher and is not in the scope of the paper. This paper shows how the rubric assists in the selection process, in that choosing topics which have high scores in each area allows for the core topics to be built out from the chosen materials. For a real course, this idea would further go into the book chosen, the pages to cover, and specific materials related to the topics.

**Week 1/2: Basic Networking / Hacker Profiles**

**Administration**: Developing basic network plans (layouts, list of materials, physical locations), understanding hackers.

**Hacking**: Vulnerabilities with physical access / layouts. How to circumvent common network setups, how hackers would influence systems based upon their profile

**Defense**: Securing systems, intrusion prevention (physical), how to prepare for possible hacking.

**Week 3: Wireless Security**

**Administration**: Building from the previous week, adding wireless to existing systems,

and developing policy for the access to the systems.

**Hacking**: How to circumvent common wireless setups, basic cracking of encryption, Warwalking/Driving.

**Defense**: Properly securing systems, understating how to best prevent attacks on a secured network, physical placement of hardware to prevent hacking.

**Week 4: Risk Assessment**

**Administration**: Develop and build a system/business policy regarding security for a business, with focus to analyzing risks by outside influence on networks.

**Hacking**: Investigate how businesses without strong risk assessment policies are vulnerable to hacking attacks both physical and digital attacks.

**Defense**: Investigate case studies of breaches in security of computers, and analyze and relate to the policy created by students.

**Week 5: Disaster Recovery**

**Administration**: Build a disaster recovery plan, with care taken to digital and non digital. Develop a backup and recovery plan for a business.

**Hacking**: Investigate how inadequate plans can leave businesses open to attacks; poorly implemented backup plans leave data open to theft.

**Defense**: Analyze disaster plans for flaws and vulnerabilities that open companies to attack.

**Week 6/7: Social Engineering**

**Administration**: continue to build on risk assessment plan to include text regarding social engineering, how to spot, and deal with attacks on people.

**Hacking**: Teach the idea behind social engineering; involve students in this aspect by having activity with a social engineering goal to achieve as class activity. Teach the fundamentals and points of access.

**Defense**: Learn how to spot social engineering attacks and how to deal with them. Understand the entry ways used most commonly by investigating case studies.

**Week 8/9: System Vulnerabilities**

**Administration**: Develop plans to quickly address security issues with systems, look into means to protect systems behind networking firewalls and equipment to prevent attacks on non-fixed systems

**Hacking**: Learn where to look for system vulnerabilities on the web, understand entry points of attacks (web, database, OS, etc). If able to, include labs to practice these attacks.

**Defense**: Learn how to harden systems against attack, develop means to quickly and efficiently determine what vulnerabilities affect systems and how to quickly patch. Understand how to spot hacked systems.

**Week 10/11: Password Cracking**

**Administration**: Develop and add to plans to include password policies, and why policies should exist regarding strong passwords. Investigate multifactor authentication and why such systems are strong.

**Hacking**: Investigate simple password cracking applications. Understand how people make passwords weak, and how to exploit for passwords using non technical means (Social engineering)

**Defense**: Setup systems help prevent password cracking, investigate password strength, and how to integrate multifactor authentication into existing systems.

**Week 12: Encryption**

**Administration**: Learn how encryption is used on systems, how and why such protection is needed. How to address systems outside of network with communications.

**Hacking**: Breaking encryption and how to try to find out encryption keys, via digital means (Brute forcing) and non digital means.

**Defense**: Investigate encryption systems, setup machines and allow for attempts to break encryptions, learn about encryption length to strength.

**Week 13/14: Proxies / Basic Network Attacks**

**Administration:** intrusion detection systems, outside access policies for systems. System logging.

**Hacking**: hiding behind proxies, how to exploit weak parts of systems, both human and non human. Use labs to attempt to break systems from various angles.

**Defense**: Understanding access points of weakened systems, how to secure hardware and software.

The class list above is just a sample of how the course ideas taken from the rubric can be molded into a course. The extra weeks at the end of the course can be weaved into the semester to include additional labs or activities to best reinforce ideas, and to allow for additional student teaching if needed. The rubric allows for the ideas to be screened and then expanded to best fit the core topics that the teacher wishes to teach. The above example shows how the main topics of Administration, Hacking, and Defense could be weaved into a course which has aspects of the ten main topics chosen based upon the rubric. This analysis allows for a set of topics which have strong cores which can be used for multiple semesters, with activities built from them as materials change.

This rubric can be applied to materials such as INFOSEC to develop a strong network security course without long periods of research, by even taking a subset of topics from a general list like INFOSEC and then using those ideas as the teaching core (Such as Administration, Hacking, and Defense was for this "course"). The rubric can be used to help in building material onto the skeleton core topics in a quick and efficient manner.

The described method above would work in any instance of teaching NS, not just in smaller schools; however the created course above was developed with these constraints because of the initial need for smaller schools to have means to develop effective courses in NS. This rubric shows that this end is attainable without large time or monetary investments by schools of this size. Of course the material being taught still needs to conform to the standards and monetary budgets of the schools implementing them. However, the rubric was designed so that these constraints are mapped along one dimension of the rubric, and materials which cannot be currently taught can still be ranked along other materials for possible future inclusion into the course. Being able to systematically rank material allows for faculty to gauge their courses against all available material and then select the material which makes the most sense to teach depending upon their needs and ability to update material. If the course being created is updated once every two years, the teacher can tailor his/her chosen materials to ones which rank highly on timelessness. However, if the teacher plans to address material every semester, then higher associability and lower timelessness material can be used. This rubric gives teachers a means to weigh their course options and time commitment and then provide students a strong course based on NS.

# Chapter 4: Existing Courses and Further Development of Rubric

## 4.1 Introduction

The main benefit of a rubric is that it provides a means for faculty interested in creating a security course to subjectively and objectively decide which material should be included. This rubric is best suited to smaller universities because it allows for a means to investigate possible topics in a more structured manner, as opposed to merely searching though materials and trying to form a class without a firm structure in place to build from. The rubric allows for this structure to be created.

Investigating what other teachers have done in their own course prior to developing a new course should be a consideration of the teacher creating a new course. With the created rubric, existing courses can be investigated by faculty to determine if the material presented in the courses are able to be easily integrated into the course being created. Examining other courses in terms of its simplicity of integration is fairly obvious; if the teacher cannot feasibly implement the material then integrating it has little use. The other areas, timelessness and associability would be applied as they are in developing a course normally. Investigating how easily the material can be built upon and how well the material can withstand time, allows for the creator of the new course to best choose items to integrate into his/her own course. Over time, a course needs to be updated to include newer materials; the rubric allows for the proposed materials in other courses to be viewed in a manner to decide if it is prudent to include or not. This factor is very important to smaller schools, which generally do not have the time to update materials continually.

Much like pulling material from books or internet sources, courses created by other schools can critiqued by the rubric as well. In general the rubric can be applied 1 to 1 on the material that is being presented in a course. For example, the labs and lectures can be looked at to determine if the material is sufficient for the created courses. Beyond the basics of the presented material, the rubric can be used to also look at other aspects of the course.

Critiquing courses can assist in choosing additional materials or ideas for a newly created course. However, unlike books which remain static until they are reprinted, an existing course may have been updated since it was last published. Because of this possibility, looking at an existing course must take into account other considerations which can be assigned a value along the rubric scale.

One of the first areas to look at for an existing course is the book(s) chosen for the lecture. In a field like NS, there are multiple viewpoints for subjects; therefore a course which takes multiple sources into account can allow for a greater perspective on the subjects. More sources can also help improve the timelessness of the course depending on the material. Having multiple sources would rank higher on associability than a single book because of the ability to draw multiple ideas from many sources.

A second area to look at is how well the labs and activities relate to the lecture material. A course created using the rubric, helps create a course in such a manner that the core of the course is easily relatable to the labs and activities, to allow for an ease of updating over time. This area of investigation would fit into all three aspects of the rubric. Simplicity is applied like it was in Chapter 3. Associability should be viewed not as much how the material relates to other pieces, but how the lab relates the lecture. Is there an easy relation from the lab to the lecture? Or is week 3's lab unrelated to week 3's lecture? Timelessness would be applied by viewing the labs contents versus the lecture, and if those ideas would need to be updated over time or if they would need to be changed to be integrated into the new course.

Finally, looking at how the lecture and lab relate to the core teaching ideas of the course allows for determining if the course should be used in some fashion. In the sample course in Chapter 3, there were three core ideas (Administration, Hacking, and Defense) and each of these three ideas was used in each chosen topic. With those three main ideas always being related to materials to teach, it allows for an ease of upgrading materials over time because the core area being taught is obvious, and easier to add to if needed. Looking at how well the materials in the course relate to their proposed core shows if the course is built to be changed over time as materials become outdated, or if the entire underlying structure of the course is based on a non-timeless idea.

Of course it is not expected that a course proposed by another teacher would be used in its entirety. However it is important to determine how the teacher designed his/her course to gain an insight into what materials could be useful to integrate into a created course. The three additional criteria above, combined with the rubric's items for critiquing existing course material (Rubric+3 [Source selection, Lecture to labs, and Objectives to teaching material]), help analyze courses from others' work, and help chose material for a new course. Simply, the +3 aspect of the rubric uses the same criteria (TAS) but employs a different view on material which makes it better suited for analyzing others' courses.

## 4.2 Applying Rubric to Existing Courses

The rubric (Timelessness, Associability, and Simplicity) along with the additional criteria (Source selection, Lecture to labs, and Objectives to teaching material) provides a good overall picture of what material could be useful to include into a newly created course. Using material from others' proposed courses provides a viewpoint into how others address the needs of teaching certain materials and which materials are looked at in greater depth. The main aspect which is important to remember when trying to augment the created course with exiting courses is the simplicity of integration. Applying the simplicity rubric idea to the course's material helps determine if it is even feasible to apply the ideas presented. For example, O'Leary's course is more of a capstone of a curriculum of NS courses. Undertaking a strictly lab based course may not be feasible to try to undertake unless it is known that the students taking the course are properly taught the basics (O'Leary, 2006). This determination must be made on a case by case basis depending upon the resources and abilities of the teacher preparing the course. Again, simplicity is a key factor in pulling material from other courses just like it was in Chapter 3.

When investigating the courses and applying the rubric+3 ideas, it should be understood that courses created by others may not fit nicely into the rubric format. O'Leary for example, teaches a course based upon case studies and labs, not a strictly lecture and lab format. Teaching a course like this does not make the course unusable or a bad resource, it simply shows how NS can be addressed from multiple angles (O'Leary, 2006). In the case of O'Leary's course, there may not be a simple link between lecture and lab, and unless the lab is laid out in detail it may be hard to use those particular topics in a new course because there is really no more basis for a topic than a simple listing of the idea itself.

To apply the rubric+3 ideas to existing courses, a similar format to Chapter 3 can be taken to help lay out what material may be useful for the course being created. Below is a sample of how this process could be done using the same course basis as Chapter 3.

The articles listed in Table 6 were found based upon searching for smaller university NS courses which is the main limiting factor for the created course. In general these courses have staff that is new to teaching NS, or have additional course loads prohibiting them from teaching only courses in NS.

**Table 6: Articles in a Similar Teaching Environment**

| School | Title | Author(s) | |
|---|---|---|---|
| Townson University | A Laboratory Based Capstone Course in Computer Security for Undergraduates | Mike O'Leary | (O'Leary, 2006) |
| Marshall University Graduate College | Teaching Students to Hack: Curriculum Issues in Information Security | Patricia Y. Logan Ph.D.<br>Allen Clarkson, M.S., MCSE | (Logan & Clarkson, 2005) |
| Miami University | Teaching Principled and Practical Information Security | Laurie Werner | (Werner, 2004) |
| Plymouth State University | Teaching Computer Security at a Small College | Cathie LeBlanc<br>Evelyn Stiller | (LeBlanc & Stiller, 2004) |
| Kennesaw State University | Teaching Students to Hack: Ethical Implications in Teaching Students to Hack at the University Level | Brian A. Pashel | (Pashel, 2006) |

Of these schools, Kennesaw has the largest total student body of around 16,000, which is a medium sized school. However most of the above schools have less than 10,000 students total, with exact numbers enrolled in NS specifically, unknown. The rubric can easily be applied to larger and smaller schools; however the main focus of this work to the latter with their additional needs of a system such as this.

Upon selection of created courses as with Chapter 3, looking over the documentation with simplicity in mind should be the first objective for picking material from existing courses. Focus should be given to the course objectives, lecture materials and labs compared to the faculty's own skill set. At this point material can be looked at from the practical and actual ability to successfully teach or integrate the ideas presented. However, unlike Chapter 3 and the filtering of topics, in most cases pieces of an existing course can pass as able to be taught even if other areas are impractical to do so. For example, O'Leary's course focuses on attack and defend labs with teams of students practicing techniques in compromising other machines. The teacher's skill set may not be sufficiently adequate to prepare a group of students to perform labs like these (especially if only a minimal number of courses teach to this end). However the sections that are practiced in the labs in his class may be able to be taught in different fashions (O'Leary, 2006).

Timelessness is an important aspect for integrating existing course material into a new course. Publications could be multiple years old with materials which are no longer valid for teaching. Looking for materials which are lasting not only provides insight into whether the course found could still be taught in its current form with minor tweaking, but also provides insight into the school and if they seem to update their course often. Like other material used in Chapter 3, a timeless piece in an existing course is more abstract, or at least makes reference to the core topic in what is being taught. For example Warner's course write-up has a table of hacking tools which perform multiple functions. The links to the applications and the application versions have most likely changed multiple times (or even became obsolete), but the root application (password cracker, network scanner, etc) is listed along with the links, allowing for searches for alternative programs to be preformed.

Associability, when analyzing existing courses, also applies similarly to how other sources are handled, generally looking at the course objectives (and how easily related to the materials presented) allows for insight into how associable material is. Of course the core topics that are chosen for the created course should also be compared to the existing material to determine how well the ideas fit. The difficulty here is that not all of the found materials have solid a set of objectives presented, only the lecture materials and perhaps labs. Again O'Leary does a good job in taking the objectives and relating then to the materials, also providing a fairly associable set of ideas along with examples, such as providing an associable topic of "packet sniffers" with possible applications to fill that gap if needed (O'Leary, 2006).

Along with the three rubric topics, there should be a few additional areas addressed in understanding existing courses dealing with the subject. These ideas should be ranked along the same criteria as the rubric topics. These additional facets of choosing material, however, are only applied to existing courses, since printed books and materials would not fit under these additional ideas.

The first of these three additional criteria is the sources used in the class for teaching. This area should focus mainly on the books and direct source material used in the class to develop and support lectures and labs. This factor is important to determine if the supplemental materials provide a large enough scope to assist in covering the taught material. The materials should be ranked on the rubric in terms of the timeless and associability of the topics contained. Simplicity can be included if the book is a possible source to use in the created course. Along with the rubric, investigate how the source is used; does it supplement lectures like a traditional course, or is it used as a lab resource which needs to be changed as the labs change? Multiple books and materials allow for a more timeless course because of the non reliance on one set of material, and can also provide multiple viewpoints on a subject. Werner's course, for example, uses two textbooks, each of which provide unique yet supplemental information on each other (Werner, 2004). Her write-up, however, does not provide a link between the books and the course, although it can be assumed that the books would be used in both lecture and lab in some fashion.

The second additional area of investigation is the relation of labs to lecture material. This is an investigation of the associability of the created course and how easily the lab work can be synthesized back into the lecture and vice versa. This idea looks into how associable the course is to use, in that it shows what labs link to the corresponding lecture and what possible changes can be made over time to keep the spirit of the presented material. A general example would be lecturing on system vulnerability of Windows XP and performing a lab to show these issues. The connection back to the lecture is fairly obvious and when the course needs to be changed to address a newer OS, the new material is fairly easy to derive from the original.

A third area of investigation into an existing course is how the objectives and core elements relate to the presented material. In general, course objectives are the milestones which taught material is viewed against. However, the rubric does not take this into account. This area of viewing existing material looks at how well the existing material's objectives relate to the material. Objectives should be fairly timeless unless they are specific and focused to a specific part of the course which can change over time (material and objective). Looking at this area in terms of associability and timelessness allows for a understanding of whether the objective is 1) abstract enough to cover multiple terms of classes, and 2) relatable to all course material in an associable manner. Again pulling material for a new course from existing courses should not require vast amounts of rework for a smaller course. For a smaller school with a non dedicated staff, trying to make relations of objectives to lecture/lab may require more time than simply choosing other materials. Warner's course presents this idea best, however the lecture and lab exercises are not listed in great detail. However from some of the objectives such as "*Install and practice using the tools of a network administrator*" (Werner, 2004) lead themselves to be easily understood as to their lab and lecture function in this course. If the objective was for example "Understand network administration and its use" it leaves more ambiguity to what possible labs could entail, thus making the idea less associable.

Application of the rubric to existing courses is the same as applying the rubric to raw material. The main difference between the two is the additional ideas which should be ranked along the same rubric scale (1-5) along the dimensions shown above. The additional items when analyzing existing materials, is a set of specific areas which may require deeper inspection beyond what the rubric addresses. Like other pieces of the rubric, additional aspects can be looked into depending upon the needs of the course being created. For example, if the created course wants to focus on alternative OS's like Linux or Apple, exploring the existing materials use and application of these areas could be added to the rubric in terms of their needs. The three aspects added above are just three basic areas which should be addressed in what the author believes to be most instances of courses.

Below is an application of the rubric for the +3 items. Keeping in mind that each section (TAS) of the +3 needs to be looked at on the 1-5 scale of the Chapter 3 rubric. The chart below gives an idea of what to think about and look for when comparing the +3 items to the TAS rubric.

**Table 7: Timelessness, Associability, & Simplicity Rubric +3 Topics**

| Idea | Source Selection | Lecture To Labs | Objectives To Teaching Material |
|---|---|---|---|
| **Timelessness** – Ability to withstand the obsolescence of ideas over time. | 1-5 based upon used books and how they score on Timeless scale in Chapter 3. | 1-5 based upon the lecture's Timelessness and how labs are impacted by changes in the lecture. | 1-5 based upon the Timelessness of the objectives compared to the changing of material presented in class. |
| **Associability** – Ability to build upon or from a core (timeless) idea. | 1-5 Based upon how well books and other source material can be replaced and associated back to topics. | 1-5 based upon how well labs can be related back to the lecture which reinforces the lab. | 1-5 based upon how easily the teaching material can be related to the objectives for the course. |
| **Simplicity** – Ability for material to be taught by staff effectively and efficiently. | 1-5 based upon the ease of using the sources into the course and the cost of doing so. | 1-5 based upon how easily the lecture and labs of course can be augmented for new course use. | 1-5 based upon how easy it is to relate found objectives to create course. |

## 4.3 Expanding The Rubric

The rubric may not address every possible scenario for creating a new course in NS or other areas. This fact is not unexpected. The rubric is not designed to be an end all approach to developing new courses, simply a means to assist in the choosing of material which is capable of being effective in the environment which it was designed for (smaller schools with limited staff). If the desire to expand the rubric exists, there are a few areas which could be addressed if the need arises for additions to the rubric.

The rubric for the most part addresses its dimensions (Timelessness, Associability, and Simplicity) in multiple parts, to address a wider range of material. Also the design of the rubric takes into account the average of a section by default, for example, if a piece of material is easily able to relate to a core topic (4 or 5 on associability) but cannot be built upon further (1 or 2 associability) the average of this would be a 3 for associability. This average allows for topics to have some flaws but still be able to be worked into a course. If desired, the dual aspects of the rubric can be broken out into multiple dimensions; this could be done if some areas are deemed to be more heavily weighted than others, where the existing rubric takes all aspects at equal weight.

The current rubric is assuming that some material is drawn from raw sources (books and other material not written at a university) and would also use some existing materials to assist in the selection of topics. The rubric does lean more towards raw sources because of the nature of the NS field and how often material changes. If desired, additional criteria can be worked into the rubric to address other aspects that are important to the course being created. For example if the rubric was being applied outside NS, for say a programming course, maybe timelessness does not apply in the same manner as it does in NS. Since syntax examples are important for programming, a whole dimension of syntax could be created and placed into the rubric instead of timelessness. Really the rubric is an associable course builder in the same way that associable materials make an easy to update NS course.

As seen in Table 7, it is fairly simple to create additional criteria to include depending upon the needs of the teacher. The rubric itself is easy to change to suit the needs of the course being created. As described above, the +3 items can be applied when pulling material from other courses which already exist, however if there is no need to do so, the rubric functions fine without the additional criteria added. This modularity allows for a uniquely crafted rubric which is tailored to the particular course being created.

# Chapter 5: Conclusions

Developing a means to select, analyze and choose materials for a college level course is something which is not necessarily a new idea. Faculty have always had to select material for courses, especially ones that need to be updated on a regular basis like NS. However the needs of smaller schools trying to determine what material they can best implement and support over time is a difficult issue to address without a means to systematically weigh their choices. The created rubric allows for a baseline from which a smaller school can select materials upon the unique constraints that they possess. The rubric is not a black box of course creation; it still requires research and the knowledge base of a subject such as NS to properly create a course. Future research can perhaps study a means to provide a more automated form of material selection. The current rubric merely provides a means to better select, analyze, and choose materials based upon ideas which are in the back of all teachers' minds; is this material within the realm of my knowledge set? Is this material not going to become outdated the semester after I teach it? And, can this subject matter build out into useful lessons? All these questions are important when choosing topics to teach. Having a simple to implement scoring model to help determine what material is best to teach allows for a more structured course to be created more quickly, and thus improves the students' overall learning experience.

Currently, smaller universities do not have a large set of documents published to address the needs to implement courses in NS; however there seems to be a desire to get material like this out for others to use. Not only is the rubric an easy to use tool, which allows for schools to assemble courses off of a baseline, but the rubric also can be used to determine if the materials in others' courses are useful for addition into a newly created course. Smaller schools are not the only ones who can benefit from the rubric however; larger schools can augment the rubric to fit the needs required by them. By applying the same rubric scale plus some additional foci to some course specific material, the rubric allows for a similar determination of the skills required and the usability of material being presented by others. This standardization allows for a greater sharing of information among a set of schools which would be better able to focus on teaching students, rather than trying to determine what is feasible for them to teach.

Like the associability contained within, the rubric also is capable of being easily tailored to fit almost any set of courses and materials. Courses like NS can benefit the most due to the volatile nature of the material, but more structured courses can tailor the rubric to best fit the needs of the course and its professor. Over time there may be a need to make additions to the rubric, most notably changes to the general selection criteria based upon implementation of the existing rubric. The possible alterations presented in Chapter 4 are the most obvious changes that could be easily made to the rubric. Needs for further investigation into creating courses based upon a rubric may exist. This alteration could be done easily by a teacher looking to create a new NS course, and allowing this teacher to determine how well the ideas presented by the rubric translate into a real life course. Depending upon how the teacher is able to create and present the material would be the best example of how well the idea of using a rubric for course creation works. The ideas behind this rubric were designed to be easily being able to be followed to create a course, but also to be easily augmented to fit the needs of the teacher.